
\magnification=\magstep1

\hsize 6.0 true in
\vsize 9.0 true in
\voffset=-.5truein
\pretolerance=10000
\baselineskip=13truept

\font\tentworm=cmr10 scaled \magstep2
\font\tentwobf=cmbx10 scaled \magstep2

\font\tenonerm=cmr10 scaled \magstep1
\font\tenonebf=cmbx10 scaled \magstep1

\font\eightrm=cmr8
\font\eightit=cmti8
\font\eightbf=cmbx8
\font\eightsl=cmsl8
\font\sevensy=cmsy7
\font\sevenm=cmmi7

\font\twelverm=cmr12
\font\twelvebf=cmbx12
\def\subsection #1\par{\noindent {\bf #1} \noindent \rm}

\def\mid {\let\rm=\tenonerm \let\bf=\tenonebf \rm \bf}

\def\para{\par \vskip 12 pt}

\def\head{\let\rm=\tentworm \let\bf=\tentwobf \rm \bf}

\def\heading #1 #2\par{\centerline {\head #1} \smallskip
 \centerline {\head #2} \vskip .15 pt \rm}

\def\eight{\let\rm=\eightrm \let\it=\eightit \let\bf=\eightbf
\let\sl=\eightsl \let\sy=\sevensy \let\m=\sevenm \rm}

\def\foots{\noindent \eight \baselineskip=10 true pt \noindent \rm}
\def\sexion{\let\rm=\twelverm \let\bf=\twelvebf \rm \bf}

\def\section #1 #2\par{\vskip 20 pt \noindent {\mid #1} \enspace {\mid #2}
  \para \noindent \rm}

\def\abstract#1\par{\para \foots {\bf Abstract: \enspace}#1 \para}

\def\author#1\par{\centerline {#1} \vskip 0.1 true in \rm}

\def\abstract#1\par{\noindent {\bf Abstract: }#1 \vskip 0.5 true in \rm}

\def\sqr#1#2{{\vcenter{\vbox{\hrule height.#2pt
  \hbox {\vrule width.#2pt height#1pt \kern#1pt
  \vrule width.#2pt}
  \hrule height.#2pt}}}}

\def\n{\noindent}
\def\s{\smallskip}
\def\m{\medskip}
\def\b{\bigskip}
\def\c{\centerline}

\def\gne #1 #2{\ \vphantom{S}^{\raise-0.5pt\hbox{$\scriptstyle #1$}}_
{\raise0.5pt \hbox{$\scriptstyle #2$}}}

\def\ooo #1 #2{\vphantom{S}^{\raise-0.5pt\hbox{$\scriptstyle #1$}}_
{\raise0.5pt \hbox{$\scriptstyle #2$}}}


.tfm scaled900
\voffset=-.5truein
\vsize=9truein
\baselineskip=24pt
\hsize=6.0truein
\pageno=1
\pretolerance=10000
\def\n{\noindent}
\def\s{\smallskip}
\def\b{\bigskip}
\def\m{\medskip}
\def\c{\centerline}
\baselineskip=14pt
\line{\hfill IUCAA - 27/94}
\line{\hfill August 1994}
\baselineskip=20pt
\c{\bf\mid Characterisation of orthogonal perfect}
\c{\bf\mid fluid cosmological spacetimes}
\m
\n {\bf Naresh Dadhich$^{\ast,\dag,} $\footnote{$^1 $}{E-mail :
naresh@iucaa.ernet.in,~~ Fax : 91 (212) 335760}, L K Patel$^{\ddag}$, K S
Govinder$^{\dag,} $\footnote{$^2 $}{E-mail : govinder@ph.und.ac.za} and P G L
Leach$^{\dag,} $\footnote{$^3 $}{E-mail : leach@ph.und.ac.za}}
\b
\noindent
$^{\ast}$ Inter University Centre for Astronomy and Astrophysics, P O Box 4,
Pune 411007, India.
\s
\n $^{\dag}$ Department of Mathematics and Applied Mathematics, University of
Natal, \break Durban 4001, South Africa.
\s
\n $^{\ddag}$ Department of Mathematics, Gujarat University, Ahmerabad --
380009, India.
\b

\n{\bf Abstract }
\s
\indent We consider the general orthogonal metric separable in space and time
variables in comoving coordinates. We then characterise perfect fluid models
admitted by such a metric. It turns out that the homogeneous models can only be
either FLRW or Bianchi I while the inhomogeneous ones can only admit $G_2 $
(two mutually as well as hypersurface orthogonal spacelike Killing vectors)
isometry. The latter can possess singularities of various kinds or none. The
non-singular family is however unique and cylindrically symmetric.
\b
\b
\b
\m
\m
\m
\noindent PACS nos: 04.20Jb, 98.80Dr
\b
\vfill\eject
\baselineskip=24pt
\item{\bf 1.} {\bf Introduction }
\s
\indent It is to the credit of general relativity (GR) that the study
of the Universe as a whole has become one of the most active areas of
scientific research and it goes by the name relativistic cosmology. The models
of the Universe we consider are given by the exact solutions of Einstein's
field equations for gravitation.. The matter content of the Universe is taken
as perfect fluid. The generally accepted model is the
Friedman-Lemaitre-Robertson-Walker (FLRW) model which describes a homogeneous,
isotropic and expanding Universe. Howsoever successful this model be,  its
homogeneity and isotropy are very special properties which could by no means be
considered as generic enough for the Universe. It would hence be important to
find cosmological solutions of Einstein's equations
without imposing the conditions of homogeneity and isotropy.
\s
\indent The first step in this direction came in the form of Bianchi models
that are homogeneous but anisotropic. Inhomogeneity was the main concern of
what is known as $G_2 $ cosmologies [1-3]. By $G_2 $ we mean a spacetime that
admits a two parameter orthogonally transitive group of isometries, that is,
there exist two spacelike Killing vectors that are mutually as well as
hypersurface orthogonal. In proper coordinates the metric depends upon only one
space variable. Inhomogeneous models have been considered by some authors [4-7]
of which the one due to Senovilla [6] is the most remarkable for being
singularity free and yet having an acceptable physical behaviour. This was a
startling result for it was generally believed on the strength of the powerful
singularity theorems [8] that the occurrence of singularity is inescapable in
GR so long as reasonable physical conditions are satisfied. Here was apparently
a counterexample to the theorems.
Subsequent to the discovery it was found that the theorems became inapplicable
because the solution in question did not obey one of the assumptions;
the existence of causal compact trapped surfaces. Not only were all physical
and
and geometrical parameters are finite and regular for whole of the spacetime,
the metric was shown to be geodesically complete [9] exhibiting the absence of
a singularity of any kind.
\s
\indent All prior attempts to manage the big-bang singularity or to construct a
non-singular cosmological model involved either unphysical behaviour for
matter, like $p < 0 $, or quantum effects and new fields or modification of GR
[10,11]. Senovilla's [6] was the first exact solution of Einstein's equations
free of any kind of singularity and possessing all physically acceptable
properties. Then the question arose, was it an isolated solution or did
there exist a family of non-singular models? Ruiz and Senovilla [12] considered
the general $G_2 $ metric separable in space and time in comoving coordinates
and identified a large family of non-singular spacetimes with cylindrical
symmetry. Is cylindrical symmetry necessary for a non-singular cosmological
model?
This is a pertinent question to be addressed next.
\s
\indent In this paper we consider a general orthogonal metric separable in
space and time in comoving coordinates and examine, in all generality, the
permissible fluid models. We  show that the already identified non-singular
family [12] is unique. It turns out that the requirement of perfect fluid
imposes $G_3, G_6 $ and $G_2 $ isometrics on the spacetime; the first two are
homogeneous Bianchi I and FLRW models while the last one alone can sustain
inhomogeneity. It turns out that  inhomogeneous models could be with or without
singularity.
However, the non-singular family is unique and is cylindrically symmetric. We
can thus characterise  all the fluid models described by an orthogonal and
separable metric [13].
\s
\indent In Sec 2 we set up the field equations for a perfect fluid for an
orthogonal metric and in Sec 3 we prove a theorem characterising perfect fluid
cosmological models and establishing uniqueness of the non-singular family. We
conclude with a discussion.
\s
\item{\bf 2.} {\bf Field Equations }
\s
\n We consider the orthogonal metric,

$$ ds^2 = D dt^2 - A dx^2_1 - B dx^2_2 - C dx^2_3 \eqno (2.1) $$

\n with velocity field given by $u = \sqrt D dt $. In comoving coordinates we
assume the metric to be separable, ie $A = A(x_{\alpha}) A(t)$ etc. The
separability can invariantly be characterised by (i) $\theta_{, \alpha} =
\theta \dot u_{\alpha} $ and (ii) $\sigma/\theta $ being constant over the
3-hypersurface. This can easily be verified from the following expressions for
the kinematic parameters

$$\theta = {1 \over 2 \sqrt D} \bigg({A_0 \over A} + {B_0 \over B} + {C_0 \over
C} \bigg), \eqno (2.2) $$

$$\sigma^2 = {1 \over 36D} \bigg[\bigg({B_0 \over B} + {C_0 \over C} - {2A_0
\over A} \bigg)^2 + \bigg({C_0 \over C} + {A_0 \over A} - {2 B_0 \over B}
\bigg)^2 + \bigg({A_0 \over A} +{B_0 \over B} - {2 C_0 \over C} \bigg)^2 \bigg]
\eqno (2.3) $$

and

$$\dot u_{\alpha} = - {D_{\alpha} \over 2D}, \eqno (2.4) $$

\n where $A_0 = \partial A / \partial t $ and $D_{\alpha} = \partial D /
\partial x^{\alpha} $.
\s
\indent We note a general result arising out of the following two relations
[14],

$$ \theta_{, \alpha} = {3 \over 2}
\bigg[( \sigma_\alpha^i + \omega_\alpha^i)_{;i} - (\sigma_{\alpha i} +
w_{\alpha_i} ) \dot u^i \bigg] \eqno (2.5) $$

$$ ~~~~= \theta \dot u_{\alpha} + { 1 \over {\sqrt g_{00}}}
\bigg( \ln \sqrt{|g/g_{00}|}\bigg)_{, 0\alpha}, \eqno (2.6) $$

\n where $\theta, \sigma, \omega, \dot{u}_\alpha$ are the
kinematic parameters; expansion, shear, rotation and acceleration,
$\dot{u}_{\alpha} = u_{\alpha; i}u^i$.  Note that we have assumed
$u_i = \sqrt{g_{00}} \delta_i^0$.
\s
\indent We infer from the above relations:
\s
\noindent {\bf Lemma :} {\it In the absence of shear and vorticity, the
expansion of
fluid is constant over the 3-space orthogonal to the fluid congruence and,
further, the acceleration also vanishes when the quantity $g/g_{00}$ is a
separable function of space and time in comoving coordinates.}
\s
\noindent {\bf Corollary :} {\it For the vorticity free spacetime with
separability (as is the case for the metric (2.1)), acceleration can be
non-zero
only if shear is non-zero.}
\s
\indent According to the Raychaudhuri equation [15], in the absence of
vorticity acceleration is necessary for
halting
the collapse to avoid the singularity which in our case can only exist if shear
is
non-zero.  Thus non-singular solutions represented by the metric (2.1) will
always have to be both inhomogeneous and anisotropic.
\s
\n Einstein's field equations are

$$R_{ik} - {1 \over 2} R g_{ik} = - 8 \pi T_{ik}, \eqno (2.7) $$

\n where for a perfect fluid

$$ T_{ik} = (\rho + p) u_i u_k - p g_{ik} \eqno (2.8) $$

\s
\indent The  explicit expressions for $T_i^k$ [16] look quite formidable and
rather intimidating.  Fortunately, we have discovered an underlying order in
them that allows us to write the rest of them from a given two (one each of
diagonal and off diagonal) by
prescribing the appropriate permutation rules.
We begin with

$$\eqalign{ -32\pi AT_0^1 &= - 2 \bigg({B_0 \over B} +
{C_0 \over C} \bigg)_1 + {A_0 \over A} \bigg({B_1 \over B} +
{C_1 \over C} \bigg) + {B_0 \over B} \bigg(-{B_1 \over B} + {D_1 \over D}\bigg)
\cr
&+ {C_0 \over C} \bigg(- {C_1 \over C} + {D_1 \over D} \bigg) \cr} \eqno (2.9)
$$

$$\eqalign{ -32 \pi T^1_1 &= {1 \over A} \bigg[{B_1 C_1 \over BC} + {D_1 \over
D} \bigg({B_1 \over B} + {C_1 \over C} \bigg) \bigg] \cr
&+ {1 \over B} \bigg[ 2 \bigg({C_2 \over C} + {D_2 \over D} \bigg)_2 + {C_2
\over C} \bigg( - {B_2 \over B} + {C_2 \over C} \bigg) + {D_2 \over D}
\bigg(-{B_2 \over B} + {C_2 \over C} + {D_2 \over D} \bigg) \bigg] \cr
&+ {1 \over C} \bigg[ 2 \bigg({B_3 \over B} + {D_3 \over D} \bigg)_3 + {B_3
\over B} \bigg({B_3 \over B} - {C_3 \over C} \bigg) \cr
&+ {D_3 \over D} \bigg({B_3 \over B} - {C_3 \over C} \bigg) + {D_3 \over D}
\bigg({B_3 \over B} - {C_3 \over C} + {D_3 \over D} \bigg) \bigg] \cr
&+ {1 \over D} \bigg[-2\bigg( {B_0 \over B} + {C_0 \over C} \bigg)_0 - {B_0
\over B} \bigg({B_0 \over B} + {C_0 \over C} - {D_0 \over D} \bigg) - {C_0
\over C} \bigg({C_0 \over C} - {D_0 \over D} \bigg) \bigg] \cr}, \eqno (2.10)
$$

\n where a subscript denotes partial
differentiation and here the assumption of separability is not effected.
\s
\indent The successive cyclic permutations $A \rightarrow B \rightarrow C
\rightarrow A $ and $1 \rightarrow 2 \rightarrow 3 \rightarrow 1 $ will give~
$T^2_0, T^3_0 $ ~from ~$T^1_0~; ~ T^2_3, T^3_1 $~ from~ $T^1_2 $;  and~
$T^2_2, T^3_3 $~ from $T^1_1 $. To write $T^1_2 $ from $T^1_0 $, let
$0 \rightarrow i2 $ (i.e. $A_0 \rightarrow iA_2, T^1_0 \rightarrow i T^1_2 $)
and $B \rightarrow C \rightarrow D \rightarrow B $ while $T^0_0 $
follows from $T^1_1 $ for  $2 \rightarrow 3 \rightarrow 1 \rightarrow i0
\rightarrow -2  (T^1_1 \rightarrow T^0_0 $) and $A \rightarrow D  \rightarrow B
\rightarrow C \rightarrow A $. Thus we can write all ten $T_i^k $, given the
two, one each of diagonal and off diagonal.
\s
\item{\bf 3.} {\bf Characterisation and Uniqueness }
\s
\indent The conditions implied by the perfect fluid character of the source
are: $T_{\alpha 0} = 0, T_{\alpha \beta} = 0 $ for $\alpha \not= \beta $ and
$T^1_1 = T^2_2 = T^3_3 $. We shall a priori assume no isometries of any kind
except the separability of the metric (2.1) in comoving coordinates. The
implementation of the fluid conditions will lead to $G_3 $ (homogeneity) and
$G_6 $ (both homogeneity and isotropy) symmetries for homogeneous models and
$G_2 $ (only admitting two spacelike Killing vectors) symmetry for
inhomogeneous models with or without singularity. Then all the fluid models
described by the metric (2.1) are characterised and uniqueness of the already
identified family of non-singular models is demonstrated. We prove the
following theorem [13].
\s
\n{\bf Theorem :} {\it The separable metric (2.1) can only represent the
following kinds of perfect fluid cosmological models :
\s
\item{(a)} if homogeneous, then Bianchi I and FLRW models,
\s
\item{(b)} if inhomogeneous, then models with or without singularity.

\n Further the non-singular family as already identified [12] is unique.}
\s
\n{\bf Proof :} Let us first of all write the three representative equations;
$T_{10} = 0, T_{12} = 0 $ and $T_1^1 = T^2_2 $,

$$ {A_0 \over A} \bigg({B_1 \over B} + {C_1 \over C} \bigg) +
{B_0 \over B} \bigg( - {B_1 \over B} + {D_1 \over D} \bigg) + {C_0 \over C}
\bigg(-{C_1 \over C} + {D_1 \over D} \bigg) = 0, \eqno (3.1) $$

$$ -2 \bigg( {C_1 \over C} + {D_1 \over D} \bigg)_2 + {B_1 \over B} \bigg({C_2
\over C} + {D_2 \over D} \bigg) + {C_1 \over C} \bigg(- {C_2 \over C} + {A_2
\over A} \bigg) + {D_1 \over D} \bigg(- {D_2 \over D} + {A_2 \over A} \bigg) =
0, \eqno (3.2) $$

$$\eqalign{&{1 \over A} \bigg[-2 \bigg({C_1 \over C} + {D_1 \over D} \bigg)_1 +
{C_1 \over C} \bigg({A_1 \over A} + {B_1 \over B} - {C_1 \over C} \bigg) + {D_1
\over D} \bigg({A_1 \over A} + {B_1 \over B} - {D_1 \over D} \bigg) \bigg] \cr
&+ {1 \over B} \bigg[2 \bigg({C_2 \over C} + {D_2 \over D} \bigg)_2 - {C_2
\over C} \bigg({A_2 \over A} + {B_2 \over B} - {C_2 \over C} \bigg) + {D_2
\over D} \bigg({A_2 \over A} + {B_2 \over B} - {D_2 \over D} \bigg) \bigg] \cr
&+{1 \over C} \bigg[2\bigg({B_3 \over B} + {A_3 \over A} \bigg)_3 -
{A_3 \over A} \bigg({A_3 \over A} - {C_3 \over C} + {D_3 \over D } \bigg) +
{B_3 \over B} \bigg({B_3 \over B} - {C_3 \over C} + {D_3 \over D} \bigg) \bigg]
\cr
&+{1 \over D} \bigg[2\bigg({A_0 \over A} - {B_0 \over B} \bigg)_0 - {A_0 \over
A} \bigg({A_0 \over A} - {C_0 \over C} + {D_0 \over D } \bigg) - {B_0 \over B}
\bigg({B_0 \over B} - {C_0 \over C} + {D_0 \over D} \bigg) \bigg]  = 0. \cr}
\eqno (3.3) $$

\s
\n{\it I}~~{\it  No isometry :} We assume no isometry to begin with. From eqn.
(3.1), $T_{\alpha 0} = 0 $, will imply

$$ {D_1 \over D} = n_1 C_1, ~{D_2 \over D} = n_1 C_2, ~{D_3 \over D} = {n_1
\over k_1} B_3, \eqno (3.4) $$

\n where

$$P_1 + k_1 Q_1 = n_1,~ B_1 = k_1 C_1 \eqno (3.5) $$

\n and

$$P_1 = {C_0/C - A_0/A \over B_0/B + C_0/C}, ~ ~~Q_1 = {B_0/B - A_0/A \over
B_0/B + C_0/C}, \eqno (3.6) $$

\n others follow by the cyclic permertation. Here $n_{\alpha} $ and $k_{\alpha}
$ are constants.
\s
\indent We integrate the exact differential
$$d(ln D) = (ln D)_1 dx_1 + (ln D)_2 dx_2 + (ln D)_3 dx_3 \eqno (3.7) $$

\n along two different paths to give

$$D(x_{\alpha}) = C^{n_1} (x_{\alpha}),~ A = C^{1/k_2} (x_{\alpha}) ~~{\rm and}
{}~B = C^{k_1} (x_{\alpha}), \eqno (3.8) $$

\n that is, the space dependence of the metric is all but determined. It
remains to find $C(x_{\alpha})$. Further for the time dependence we get

$$k_2 (1 + k_1) {A_0 \over A} + (1 + k_2) {B_0 \over B} + (1 + k_1 k_2) {C_0
\over C} = 0 \eqno (3.9) $$

\n and $n_1 = 1 + k_1 + 1/k_2 $.
\s
\indent It may be noted that we have so far used only the three equations
$T_{\alpha 0} = 0 $ to obtain the relations (3.8) and (3.9) which leave only
$C(x_{\alpha}) $ and two of $A(t), B(t), C(t) $ to be determined. Substituting
(3.8) in (3.2) and its permutants leads to $C(x_{\alpha}) = const. $. Thus the
metric can only represent a homogeneous Bianchi I model.
\s
\indent On the other hand when $A_0/A = B_0/B = C_0/C $, the shear vanishes and
so does the acceleration. The spacetime is then both homogeneous and isotropic
which determine FLRW uniquely [17]. When $A_0/A = B_0/B \not= C_0/C $, eqn.
(3.1) and its permutants will imply either Bianchi I or the spacetime admits a
$G_1 $ isometry. This is the case we consider next.
\s
\n{\it II}~~ {\it $G_1 $ isometry :}  Let $\partial/\partial x_3 $ be the
spacelike Killing vector and hence the metric is a function of only two space
variables, $x_1 $ and $x_2 $.
\vfill\eject
\indent Note that eqn. (3.3) has the form

$$ {f_1 \over A(t)} + {f_2 \over B(t)} + {f_3 \over C(t)} = F(t) \eqno (3.10)
$$

\n where $f_1, f_2, f_3 $ are functions of $x_{\alpha} $,  containing
respectively derivatives with respect to $x_1, x_2 $ and $x_3 $. In this case
$f_3 = 0 $ and $T_{30} \equiv 0 $. Eqn. (3.10) gives rise to two cases: (i)
$A_0/A = B_0/B \not= C_0/C $ and  (ii) $ A_0/A \not= B_0/B \not= C_0/C $ and
$f_1 = const.,~ f_2 = const. $. That means $\sigma $ is non-zero to give
non-zero acceleration. It could be a viable case for a non-singular model as
well.
In (i) we can set $A = B $ for $A(t) = B(t) $ is implied by $A_0/A = B_0/B $ (a
constant multiple can always be absorbed) and $A(x_{\alpha}) = B(x_{\alpha}) $
can be done by an appropriate coordinate transformation. Eqn. (3.1) implies
$C(x_{\alpha}) = D^{\lambda}(x_{\alpha}) $ and eqns.
(3.1) -- (3.3) give three equations to determine the space dependence of the
metric. The Lie group analysis of the equations (see Appendix) leads to the
inference that the functional dependence can only occur in the form $A (x_1 +
x_2), A(x^2_1 + x^2_2) $ and $A(x_1/x_2) $. The first two cases reduce to
single variable dependence by suitable coordinate transformation, which we
consider separately. The last case is obviously singular and could not be
considered as a viable case for any kind of cosmology.
\s
\indent In (ii) $A_0/A \not= B_0/B \not= C_0/C $, following the same route we
get from $T_{\alpha 0} = 0; A, B$ and $ D $ in terms of $C(x_{\alpha}) $ as
before. Then $T_{12} = 0 $ determines $C(x_{\alpha}) = (f(x_1) + f(x_2))^c $.
Eqn. (3.10) represents two equations;  $f_1 = const., ~f_2 = const. $ and two
more similar equations. These will ultimately determine $C(x_{\alpha}) = const.
$ and again the spacetime is Bianchi I.
\s
\indent Thus $G_1 $ symmetry does not yield a viable fluid model.
\s
\n{\it III}~~{\it $G_2 $ isometry:} Finally we have the spacetime general
enough to sustain an inhomogeneous fluid. All fluid models are inhomogeneous
and anisotropic. In view of eqns. (2.5) and (2.6), it follows that
inhomogeneous spacetime has necessarily to be anisotropic. Inhomogeneous models
can have singularities of different kinds or none.
\s
\n Ruiz and Senovilla [12] have thoroughly analysed this case and have shown
that the spacetime possesses a rich singularity structure. The metric (2.1)
will have models with singularity but not always of the big-bang kind as well
as models free of singularity. The latter family is shown to be unique and
cylindrically symmetric. Since non-singular solutions
 are allowed only in this case, the identified non-singular family is unique
for the general orthogonal metric (2.1).
\s
\indent This completes the proof of the theorem.
\s
\indent The most general non-singular metric [12] is given by,

$$\eqalign{ds^2 &= \cosh^{1+n} (at) \cosh^{n-1} (nar) \bigg( dt^2 - {\sinh^2
(nar) \over P^2} dr^2 \bigg) \cr
&- \cosh^{1+n} (at) {P^2 \over n^2 a^2 L^2 \cosh^{{n-1 \over n}} (nar)} d
\phi^2 - {\cosh^{1-n} (at) \over \cosh^{{n-1 \over n}} (nar)} dz^2 \cr}, \eqno
(3.11) $$

\n where

$$L = K - {K-1 \over 2n},~ P^2 = \cosh^2 (nar) + (K-1) \cosh^{{2n-1 \over n}}
(nar) - K \eqno (3.12) $$
\vfill\eject
\n and $K, n, a $ are constants. The coordinates range as $- \infty < t, z <
\infty,~0 \leq r < \infty,~0 \leq \phi \leq 2 \pi $ and the metric has
cylindrical symmetry.
\s
\indent The fluid parameters are given by

$$8 \pi \rho = X \bigg[ (n-1) (2n-1) (n+3) K \cosh^{-2} (nar) + (n+1) (n-3)
\cosh^{-2} (at) \bigg] \eqno (3.13) $$

$$8 \pi p = X \bigg[ (n-1)^2 (2n-1) K \cosh^{-2} (nar) + (n+1) (n-3) \cosh^{-2}
(at) \bigg], \eqno (3.14) $$

\n where

$$X = {a^2 \over 4} \cosh^{-(1+n)} (at) \cosh^{1-n} (nar). \eqno (3.15) $$

\n Both density and pressure are positive and $p \leq \rho $~for ~$K \geq 0 $.
The equation of state $\rho = 3p $ for radiation  is admitted
when $n=3 $ .  Senovilla's model [6] further requires $K = 1 $. The case $K = 0
$ gives the stiff fluid equation of state, $\rho = p $ [18]. The case $K = 1 $
has been considered separately and it has been shown that radial heat flux can
be incorporated without disturbing the singularity--free character of the
metric [18,19].
\s
\item{\bf 4.} {\bf Discussion }
\s
\indent The main result of the paper is that the already identified family of
non-singular cosmological models is unique not only for the $G_2 $ metric but
also for the general orthogonal metric separable in space and time in comoving
coordinates. Thus the complete set of non-singular solutions has been
identified. In the process of establishing this result we have also been able
to characterise all perfect fluid models described by the metric (2.1). They
are: homogeneous Bianchi I and FLRW; and inhomogeneous with or without
singularity (different kinds of singularities occur [12]). We assume no
isometries a priori, the perfect fluid conditions imply $G_3 $ and $G_6 $
symmetries for homogeneous, and $G_2 $ for inhomogeneous models.
\s
\indent The non--singular character of fluid models singles out cylindrical
symmetry. Like inhomogeneity, it is only a necessary condition but not
sufficient. A kind of formal connection can be indicated between the
non--singular metric (3.11) with $K = 1 $ and the FLRW open model. The former
can be thought of as arising out of a natural inhomogenisation of the latter
[20]. The unfortunate feature of the metric (3.11) is that anisotropy does not
decay with time (since $\sigma/\theta = {\rm const} $), which means it can
never evolve into FLRW. There may, however, occur a non--singular solution when
the assumption of separability is dropped, which may isotropise to FLRW at late
times. That would be a very significant result for cosmology, but the situation
becomes mathematically formidable.  We are currently investigating this
question for spherical and cylindrical symmetry.
\s

\n{\bf Acknowledgements}
\s
\n ND thanks A K Raychaudhuri for constructive criticism and discussions and
the
University of Natal for the award of a Hanno Rund Fellowship.  LKP thanks IUCAA
for hospitality while part of this work was done.  KSG and PGLL thank the
University of Natal and the Foundation for Research Development of South Africa
for their continuing support.

\vfill\eject

\n {\bf Appendix}
\s
\n An $n$th order system of differential equations

$$ {\bf E} ({\bf x}, {\bf y}, {\bf y}',\ldots,{\bf y}^{(n)}) = 0  \eqno (A.1)
$$

\n will have the Lie (point) symmetry [21]

$$ G = \xi_i(x_i, y_i) {\partial \over \partial x_i} + \eta_i (x_i, y_i)
 {\partial \over \partial y_i} \eqno (A.2) $$
\n iff
$$ G^{[n]} {\bf E}_{\bigg|_{_{\bf E = 0}}} = 0, \eqno (A.3) $$

\n where $G^{[n]}$ is the $n$th extension of $G$ needed to take care of the
$n$th derivatives in (A.1) .
\s
\n We analyse the following three equations:

$$2 \bigg({D_1 \over D} \bigg)_2 - (\alpha - 2) {D_1 D_2 \over D^2} - {D_1 A_2
+ D_2 A_1 \over D A} = 0, \eqno (A.4) $$

$$ 2 \bigg({D_1 \over D} \bigg)_1 - (\alpha - 2) {D^2_1 \over D^2} - {2 D_1 A_1
\over DA} = 2 \bigg({D_2 \over D} \bigg)_2 - (\alpha - 2) {D^2_2 \over D^2} -
{2 D_2 A_2 \over DA}, \eqno (A.5) $$

$$\eqalign{&{D \over A} \bigg[ 2 \lambda \bigg({D_1 \over D} \bigg)_1 + \lambda
(\lambda + 1) {D^2_1 \over D^2} - 2\bigg({A_1 \over A} \bigg)_1 - (\lambda + 1)
{A_1 D_1 \over AD} \bigg] \cr
&+ {D \over B} \bigg[- 2 \bigg({A_2 \over A} \bigg)_2 - 2 \bigg({D_2 \over D}
\bigg)_2 + (\lambda - 1) {D^2_2 \over D^2} + (\lambda + 1) {A_2 D_2 \over AD}
\bigg] = l \cr}, \eqno (A.6) $$

\n where $\alpha = (1 + 2 \lambda - \lambda^2)/(1 + \lambda) $ and $l$ and
$\lambda$ are constants.
\s
\indent The standard Lie analysis [22] gives the Lie point symmetries of
the above equations  as
$$ G_1 = {\partial \over \partial x_1} \eqno (A.7) $$

$$G_2 = {\partial \over \partial x_2} \eqno (A.8) $$

$$G_3 = x_1 {\partial \over \partial x_1} + x_2 {\partial \over \partial x_2} -
2A {\partial \over \partial A} \eqno (A.9) $$

$$G_4 = A {\partial \over \partial A} + D {\partial \over \partial D} \eqno
(A.10). $$

\n To reduce the system of partial differential equations
(A.4)--(A.6) to one of ordinary differential equations
(and hence solve them) we need to decide on an appropriate
independent variable.
{}From (A.7) and (A.8) we have the possiblity of independent variables
$$ u = c_1 x_1+ c_2 x_2  \eqno (A.11) $$

\n and from (A.9)

$$u = {x_2 \over x_1}. \eqno (A.12) $$

\n The fourth symmetry implies that $u = u(x_1,x_2)$.
\s
Now, (A.11)  implies

$$ A = A(x_1+x_2) ~~~D = D(x_1+x_2) \eqno (A.13) $$

\n while (A.12) implies

$$ A = {1 \over x_1} {\cal A} \bigg({x_2 \over x_1} \bigg) \eqno (A.14) $$

\n with
$$ D = x_1 {\cal D} \bigg({x_2 \over x_1} \bigg) \eqno (A.15) $$

\n coming from the addition of (A.10). On the other hand, $u=u(x_1,x_2)$
implies

$$ A=A(u), ~~ D = D(u) \eqno (A.16) $$
\n which further implies

$$ u=f(x^2_1 + x^2_2).  \eqno (A.17) $$
\s
\indent We could take other combinations of the symmetries.  The only one of
relevance
is (A.9) + $k $ (A.10).  This gives
$$ A = x^{k-2}_1 f \bigg({x_2 \over x_1} \bigg), ~~D = x^k_1 \bigg({ x_2 \over
x_1} \bigg) \eqno (A.18) $$

\n and is equivalent to (A.14) and (A.15) for the purposes of determining
whether $A$ and $D$ contain singularities.

\vfill\eject
\n{\bf References}
\s

\item{[1]} C.G. Hewitt and J. Wainwright, Class. Quantum Grav. 7, 2295 (1990).
\s
\item{[2]} J. Wainwright, J. Phys. A {\bf 14}, 1131 (1981).
\s
\item{[3]} M. Carmeli, Ch. Charach, and S. Malin, Phys. Rep. {\bf 76}, 80
(1981).
\s
\item{[4]} J. Wainwright and S.W. Goode, Phys. Rev. {\bf D 22}, 1906 (1980).
\s
\item{[5]} A. Feinstein and J.M.M. Senovilla, Class. Quantum Grav. {\bf 6}, L89
(1989).
\s
\item{[6]} J.M.M. Senovilla, Phys. Rev. Lett. {\bf 64}, 2219 (1990).
\s
\item{[7]} W. Davidson, J. Math. Phys. {\bf 32}, 1560 (1991).
\s
\item{[8]} S.W. Hawking and G.F.R. Ellis, The Large Scale Structure of the
Universe, Cambridge University Press, 1973.
\s
\item{[9]} F.J. Chinea, L. Fernandez-Jambrina, and J.M.M. Senovilla, Phys. Rev.
{\bf D 45}, 481 (1992).
\s
\item{[10]} J.M. Murphy, Phys. Rev., {\bf D 8}, 4231 (1973).
\s
\item{[11]} D. Berkenstein and A. Meisels, Ap. J., {\bf 237}, 342 (1980).
\s
\item{[12]} E. Ruiz and J.M.M. Senovilla, Phys. Rev. {\bf
D45}, 1995 (1992).
\s
\item{[13]} N. Dadhich and L.K. Patel, Preprint IUCAA 14/94, July 1994.
\s
\item{[14]} A.K. Raychaudhuri, Private communication (1993).
\s
\item{[15]} A.K. Raychaudhuri, Phys. Rev. {\bf 90}, 1123 (1955).
\s
\item{[16]} R.C. Tolman,
Relativity, Thermodynamics and Cosmology (Oxford, 1934).
\s
\item{[17]} S. Weinberg, Gravitation and Cosmology (John Wiley, 1972).
\s
\item{[18]} N. Dadhich, L.K. Patel and R.S. Tikekar, Preprint: IUCAA 19/93.
\s
\item{[19]} L.K. Patel and N. Dadhich, Class. Quantum Grav. {\bf 10}, L 85
(1993).
\s
\item{[20]} N. Dadhich, R. Tikekar and L.K. Patel, Current Science {\bf 65},
694 (1993).
\s
\item{[21]} P.J. Olver, Applications of Lie Groups to Differential Equations
(Springer--Verlag, 1993).
\s
\item{[22]} A.K. Head, Comp. Phys. Comm. {\bf 77}, 241 (1993).

\bye